\documentclass[12pt,a4paper,final]{iopart}

\usepackage{graphicx}
\usepackage[breaklinks=true,colorlinks=true,linkcolor=blue,urlcolor=blue,citecolor=blue]{hyperref}

\expandafter\let\csname equation*\endcsname\relax
\expandafter\let\csname endequation*\endcsname\relax
\usepackage{amsmath}
\usepackage{amssymb}

\begin{document}

\title[Machine learning in sentiment
reconstruction of the simulated stock market]
{Machine learning in sentiment
reconstruction of the simulated stock market}

\author{Mikhail Goykhman}
\address{Enrico Fermi Institute, University of Chicago,\\
Chicago, IL 60637, USA}
\vspace{0.1cm}
\address{Racah Institute of Physics, Hebrew University of Jerusalem,\\
Jerusalem, 91904, Israel}
\ead{\mailto{goykhman@uchicago.edu} \footnote{\mailto{goykhman89@gmail.com}}}

\author{Ali Teimouri}
\address{Consortium for Fundamental Physics, Lancaster University,\\
 Lancaster, LA1 4YB, United Kingdom}
\ead{\mailto{a.teimouri@lancaster.ac.uk}}

\begin{abstract}

In this paper we continue the study of the simulated stock market framework defined
by the driving sentiment processes. We focus on the market environment driven
by the buy/sell trading sentiment process of the Markov chain type.
We apply the methodology of the Hidden Markov Models
and the Recurrent Neural Networks to reconstruct
the transition probabilities matrix of the Markov sentiment
process and recover the underlying sentiment states
from the observed stock price behavior.

\end{abstract}

\section{Introduction}

A typical stock market simulation framework considers a discrete-time evolution of
a system of agents, each possessing shares of
stock and units of cash, who submit orders to the
stock exchange, see 
\cite{PalmerAH1994,ArthurHL1996,PalmerAH1998,LeBarronAP1999,Raberto2000,Bonabeau2002,
PastorePC2010,PontaRC2011,Bertella2014,Goykhman2017} for some of the original papers. One can assume that
at each time step each agent participates in a trade
with some probability, which in the simplest models is identical for all of the agents, and is a
constant in time. If the agent decides to participate in a trade
then it needs to decide on the side of the trade (buy or sell),
the limit price (for which it is willing to buy or sell the shares of stock),
and the size of the order.

In the simplest models the buy/sell side is determined 
by a flip of a fair coin, the limit price is normally distributed around the value related
to the most recent stock price, and the size of the order is drawn uniformly
between a zero and all-in \cite{Raberto2000}. Under such conditions the stock price time series
will exhibit a mean-reverting behavior around the equilibrium, $P_e=M/S$,
determined by the total amount of cash $M$ and the total number of shares $S$
in the system. This relation is a simple consequence of the balance
of an expected cash flows to and from the stock market capitalization.
Similarly, the volatility $\sigma_e$ around the mean price $P_e$ is determined by the standard deviation $\sigma$
of the limit orders
submitted by the agents, $\sigma_e\simeq k\sigma$, for a certain value of $k$ \cite{Raberto2000}.

The real world stock prices time series are far from being simple mean-reverting processes.
In order to obtain interesting stock price dynamics one
needs to incorporate a non-trivial behavior of the agents,
rather than a random behavior described in the previous paragraph. Various models
have been proposed in the literature, incorporating a sophisticated
strategies into the behavior of agents, such as trend-following, contrarian, fundamental,
utility optimization, etc, see \cite{SamanidouZS2007} for a review. The common feature of those models is that
the agents apply a specified strategy to the observed stock price behavior to make a purposeful decision
about their trading actions. Such models have been rather successful
in explaining stylized facts of a stock price behavior, such as fat tails of logarithmic returns \cite{Mandelbrot1963}
and volatility clustering.

Another way to model the stock price behavior in a simulated setting has been proposed in \cite{Goykhman2017}.
The starting assumption adopted in \cite{Goykhman2017} was to consider a stock market framework
in which the strategies of the agents are in one-to-one correspondence with 
a set of processes, called sentiments.
\footnote{There are subtle specifics as to how this one-to-one correspondence
is to be understood. For instance, we do not distinguish between market frameworks in which a subgroup of agents
follows a particular strategy, or each agent in the whole system adopts that
strategy with the corresponding probability.}
Examples of such sentiment processes
include the perceived volatility sentiment, determining the standard deviation of the submitted limit price,
the buy/sell attitude, determining the willingness to buy rather than sell a stock,
and the participation sentiment, determining the trading volume.
All or large groups of the agents receive sentiments from the same
source, probably with some noise around it. For instance,
half of the agents might receive an information that a stock has 
been assigned a positive rating, and therefore the agents in that group will be, {\it e.g.}, thirty-percent more
willing to buy that stock rather than to sell it.

One could ask what would be the reason to model the behavior of the stock
market participants using the driving sentiment processes.
Indeed, it is likely to expect that the agents participating in the market
will readjust their trading decision basing on the observed stock
price behavior, rather than persist following the pre-specified sentiment.
We answer this question by pointing out that the sentiment
processes which we discuss in this paper are emergent rather
than imposed, in the sense that the collective behavior of the agents, in the framework
considered in this paper, can be described via sentiment states.

Therefore the sentiment trading conditions discussed in this paper
define a market framework as the starting assumption.
In a sentiment-driven stock market framework the stock price
dynamics is largely determined by the properties of the underlying
sentiment processes. Once all the sentiments have been specified
we can predict well what will come out of the simulation. 
We can ask the opposite question: if we observe a stock
price behavior and we known that it has originated in a sentiment-driven
framework, how do we determine the underlying sentiment processes?
This paper is concerned with such a question. 

Notice that above we are talking about attempting to explain
a stock price behavior using sentiment driving processes
when we know for sure that the observed stock price time series
has been simulated in a sentiment-driven market simulation.
One can ask a question of whether the real-world stock data
can be analyzed in a similar way, starting from the assumption
that the behavior of the real market participants boils down,
within a certain degree of approximation, to a few driving
sentiment processes. In this spirit it would be interesting to explore
how the sentiment market framework can account for the real market behavior.
We will not be addressing such questions in this paper,
leaving it for future work.

\vspace{0.5cm}

The simplest sentiment-driven stock market environment is defined
by a few well-separated sentiment regimes. By calculating the mean stock
price in each of those regimes we can derive the corresponding
sentiment.
The probability to switch between the regimes is then small (of order of
an inverse number of steps the market spends in the given sentiment state).
We discuss such a situation in section~\ref{sec:buy_sell_sentiment},
where we consider the market environment in which various groups of
agents follow the buy/sell sentiment which changes twice over the time of the simulation.
We demonstrate explicitly that the resulting mean stock price in each of the
sentiment regimes is consistent with the cash flow balance equation.

A more sophisticated situation is to consider a non-trivial sentiment time series,
switching regularly between states with different sentiments.
A simple example of such a process, which we will be focusing on for the most
part of this paper, is given by a Markov chain, with a certain
transition probability matrix.
The problem
is then to recover the transition probability matrix of the sentiment Markov process
from the observed stock price time series. We will address this problem in section \ref{section_sentiment_hmm_reconstruction} using the 
Baum-Welch algorithm of the Hidden Markov Model (HMM). For an excellent
review of the HMM we refer reader to \cite{Rabiner1989}.

We notice for completeness that
there have been some
attempts in the literature to use techniques of the HMM to predict
the stock price behavior \cite{Hassan2005,Gupta2012,Kavitha2013}. A typical goal stated in the literature is
to predict the next day's stock price using the HMM trained on the most
recent stock price values. We notice that since typically the next day's stock
price is correlated with the current day's stock price, any prediction
prescription for the next day's price which relies on the current day's price
will appear to be successful, and exhibit a high correlation scores with the actual
next day's stock price.
However we point out that
one should be careful about how valuable is in fact such a prediction,
and whether one can construct a profitable trading algorithm based on it.
In the frameworks aspiring to predict the next day's stock price we suggest to test this goal by
calculating the fraction of days on which at least the direction (and at most the return) of the stock price 
has been guessed correctly. In this paper we are not concerned with an interesting question of predicting
the real world stock
prices using the HMM.

If we know the transition probabilities matrix for the sentiment Markov process
and if we know the current sentiment state which we are in, then we can
make an informed trading decision, taking into account the most likely
true market value of the stock, and the probability that such a value will change,
and by how much. In section  \ref{section_sentiment_hmm_reconstruction}
we discuss application of the Viterbi algorithm of the HMM to the problem
of reconstruction of the underlying hidden sentiment states. We demonstrate that
due to limitations of the HMM to describe the sentiment market framework
the Viterbi algorithm performs poorly, and predicts the underlying sentiment
states with an accuracy being as good as a random guess.
We explain in subsection \ref{sub_sec:reconstruction} that this is essentially
due to the fact that the distribution of observed stock prices depends not only on the current
sentiment but also on the sequence of the recent sentiments, as contrasted with
the local condition of the HMM.

Therefore in order to predict the underlying
hidden sentiment states one needs a different method, which would
be capable to remember the sequence of states for a fit, rather than a single state.
We suggest that for this purpose one can use the technology of the
Recurrent Neural Networks (RNN). In section \ref{Sentiment_fit_using_Recurrent_Neural_Network} we
demonstrate that using the RNN one can improve the prediction of the underlying sentiment
states from the observed stock prices, and make it significantly above a random guess.

 We discuss
our results in section \ref{sec:discussion}.
In \ref{sec:hmm} we review the methodology of the Hidden Markov Models
relevant for this paper. \ref{sec:rnn} is dedicated to a review of the Recurrent Neural Networks.

\section{A simple buy/sell sentiment market}
\label{sec:buy_sell_sentiment}

In this paper we study the stock market simulation framework
in which all of the non-trivial price dynamics originates
from a specified processes, which we refer to as sentiments.
In this section we discuss a simple example of a sentiment-driven
stock market system, and explain our notation on this system.
We will be only considering a non-trivial buy/sell sentiment process.
It is defined as follows: the agent $A$ is said to follow a buy/sell sentiment $\psi(t)$
if at the time $t$, provided it decides to participate in a trade,
it will submit a buy order with the probability $p_b$ and a sell order with the probability $p_s=1-p_b$,
determined as
\begin{equation}
\frac{p_b}{p_s}=e^{\psi(t)}\,.
\end{equation}

Consider the system of $N=1000$ agents, evolving in a discrete
time $t=1,\dots,T$ over the course of $T=10^4$ steps. Each agent $A_i$,
$i=1,\dots,N$, at the beginning of the considered time step, 
is in a possession of $M_i$ units of cash, and $S_i$ shares of stock.
The amount of shares of stock and the amount of cash in the system is constant.
At each time step each agent will submit a trade order
\footnote{A typical way to simulate trading between the agents which
can be found in the literature is done
by maintaining a limit order book an operating a matching engine,
similar to the real stock exchange. In this paper we clear the unfilled orders
from the order book before the next time step. We refer reader
to \cite{Goykhman2017} for a recent discussion of the mechanics of the matching engine.}
with probability $\rho=1/10$
to the stock exchange.
Regardless of the side of the order the agent will submit a limit price for its order
by taking a gaussian draw centered around the most recent stock 
price with the standard deviation $\sigma=0.01$.

\begin{figure}
\begin{center}
 \includegraphics[width=12.8cm, height=8cm]{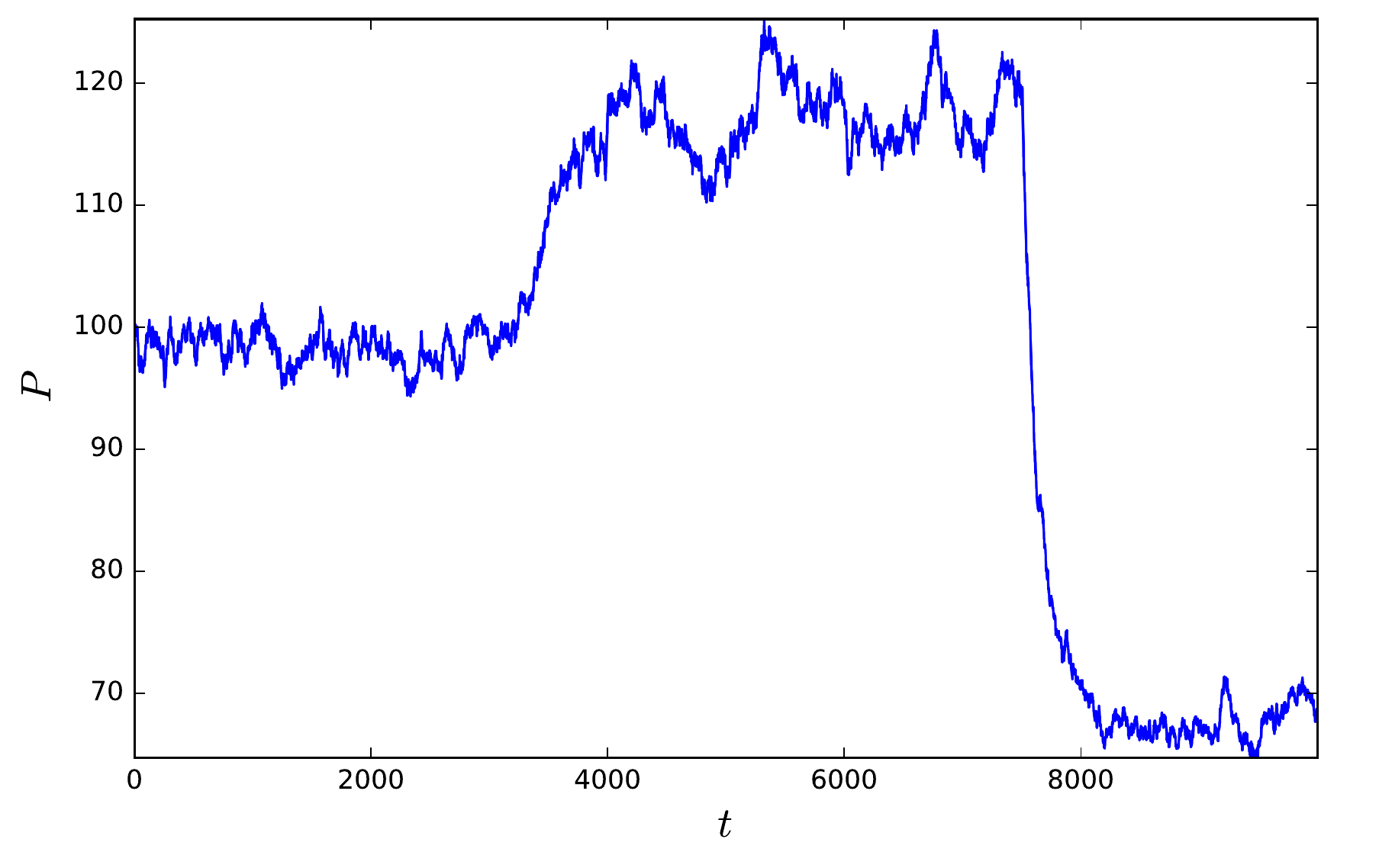}
 \caption{ \label{stocks_section_2} Stock time series for the simulation
 in section \ref{sec:buy_sell_sentiment}.
 There are three sentiment regimes: ${\cal T}_1\simeq [0,T/3]$, with
 the mean $P_e^{(1)}=98.5$ and volatility $\sigma^{(1)}=1.5$,
 ${\cal T}_2\simeq [T/3,3T/4]$, with
 the mean $P_e^{(1)}=116.7$ and volatility $\sigma^{(2)}=3.3$,
 ${\cal T}_3\simeq [3T/4,T]$, with
 the mean $P_e^{(1)}=69.3$ and volatility $\sigma^{(3)}=4.3$.
 These values are in agreement with the prediction (\ref{price_psi_simple}),
 with $P_1=100$.
}
\end{center}
\end{figure}

The side of the order (buy or sell) in the system discussed in this section is determined as follows.
The agents are divided into two groups: the group $G_1$ has $zN$ agents, $z=1/4$,
and the group $G_2$ has the remaining $(1-z)N$ agents. The agents of the group $G_1$
are following the sentiment process
\begin{equation}
\label{psi_1_simple}
\psi_1(t) = \left \{
  \begin{aligned}
    &0  && \text{if}\ t\in [0,T/3] \\
    &\log\, 2, && \text{if}\ t \in [T/3,T]
  \end{aligned} \right.
\end{equation} 
and the agents of the group $G_2$
are following the sentiment process
\begin{equation}
\label{psi_2_simple}
\psi_2(t) = \left \{
  \begin{aligned}
    &0  && \text{if}\ t\in [0,3T/4] \\
    &-\log\,2, && \text{if}\ t \in [3T/4,T]
  \end{aligned} \right.
\end{equation} 
We will assume that regardless of the side each agent $A_i$ submits an order of a uniform size $u_i={\cal U}(0,1)$,
and for a limit price ${\cal N}(P_{t-1},\sigma)$, where $P_{t-1}$ is the most recent stock price.
These conditions are designed in the spirit of the sentiment-driven market
framework of \cite{Goykhman2017}, with a constant volatility sentiment.

The starting stock price is chosen to be $P_1=100$. We give each agent $M_i(t=1)={\cal N}(10^5,10^3)$
units of cash and $S_i(t=1)=10^3$ shares of stock. Non-identical initial wealth allocations
in the sentiment-driven market framework
have been considered in \cite{Goykhman2017}. Under the conditions of a  uniformly distributed
trade size and a neutral buy/sell sentiment, the equilibrium price is $P_e=M_i/S_i=100$,
equal to the original stock price $P_1$.
Therefore we expect that until the time $t=T/3$ the stock price will be mean-reverting around the $P_1$.
After that the sentiment of the agents from the group $G_1$ becomes more bullish, so we expect
the price to go up to a new equilibrium value. The length of the transition period depends on the 
perceived stock volatility $\sigma$ and the trading intensity of the agents. 
For the purposes of this section
we skip discussion of
the transition period and  calculate the new equilibrium price of a well-separated new sentiment regime.

In equilibrium the average cash flows to and from the stock market capitalization should 
balance each other. This condition can be derived by averaging the cash flow balance equation
\begin{equation}
\label{flow_balance}
p_b^{{\rm eff}}\,M=p_s^{{\rm eff}}\,S\,P_e\,,
\end{equation}
where we have denoted the total cash in the system as $M=\sum_iM_i$, and the total number of shares
outstanding as $S=\sum_iS_i$, and introduced the effective buy and sell probabilities.
The latter are defined by noticing that having several groups
of agents with different sentiments in each group is the same as having one group, where each agent decides
at each step to act accordingly with a certain sentiment, with the corresponding probability:
\begin{align}
p_b^{{\rm eff}}&=p({\rm buy}|G_1)\,p(G_1)+p({\rm buy}|G_2)\,p(G_2)
=\frac{e^{\psi_1}}{e^{\psi_1}+1}\,z+\frac{e^{\psi_2}}{e^{\psi_2}+1}\,(1-z)\,,\\
p_s^{{\rm eff}}&=p({\rm sell}|G_1)\,p(G_1)+p({\rm sell}|G_2)\,p(G_2)
=\frac{1}{e^{\psi_1}+1}\,z+\frac{1}{e^{\psi_2}+1}\,(1-z)\,.
\end{align}

Specifically for the sentiment processes (\ref{psi_1_simple}), (\ref{psi_2_simple})
we obtain the expected equilibrium price time dependence
\begin{equation}
\label{price_psi_simple}
P_e(t) = \left \{
  \begin{aligned}
    &P_1  && \text{if}\ t\in [0,T/3] \\
    &\frac{13}{11}\,P_1  && \text{if}\ t\in [T/3,3T/4] \\
    &\frac{5}{7}\,P_1 && \text{if}\ t \in [3T/4,T]
  \end{aligned} \right.
\end{equation} 
This can be easily confirmed by a simulation, see figure \ref{stocks_section_2}.

In the setup considered in this section the market has undergone just two
transitions between three sentiment regimes, over a large number $10^4$ of simulation
steps. In such a case different sentiment regimes are well separated from
each other, and we can easily calculate the mean
stock price and the standard deviation for each sentiment regime, see caption of figure \ref{stocks_section_2},
and use the flow balance equation (\ref{flow_balance}) to calculate the corresponding sentiment.
We can also infer the order of magnitude of the transition
probabilities between the sentiment regimes as being negligibly small, $P_{trans}\simeq {\cal O}(10^{-4})$.

\section{Sentiment fit using Hidden Markov Models}
\label{section_sentiment_hmm_reconstruction}

In section \ref{sec:buy_sell_sentiment} we discussed the simplest example
of inferring sentiment properties from the observed stock price time series
in a system with a well-separated sentiment regimes.
A more sophisticated example of a sentiment-driven market can be constructed
by considering a sentiment which is itself a non-trivial time series process.
In this section we consider a simple market simulation
where trading activity of the agents is influenced by a simple
buy/sell sentiment process $\psi(t)$ of the Markov chain type.
In subsection \ref{sub_sec:simulation}
we describe specifics of the simulated stock market.
In subsection \ref{sub_sec:reconstruction} we discuss
how one can reconstruct the properties of the underlying sentiment 
process using the  framework of the HMM. In subsection \ref{repeated_hmm}
we describe the outcome of multiple simulations and the HMM fit of the sentiments.
We refer reader to \ref{sec:hmm} for a review of the HMM relevant
for this paper.

\subsection{Simulation setup}\label{sub_sec:simulation}

\begin{figure}
\begin{center}
 \includegraphics[width=14cm, height=9cm]{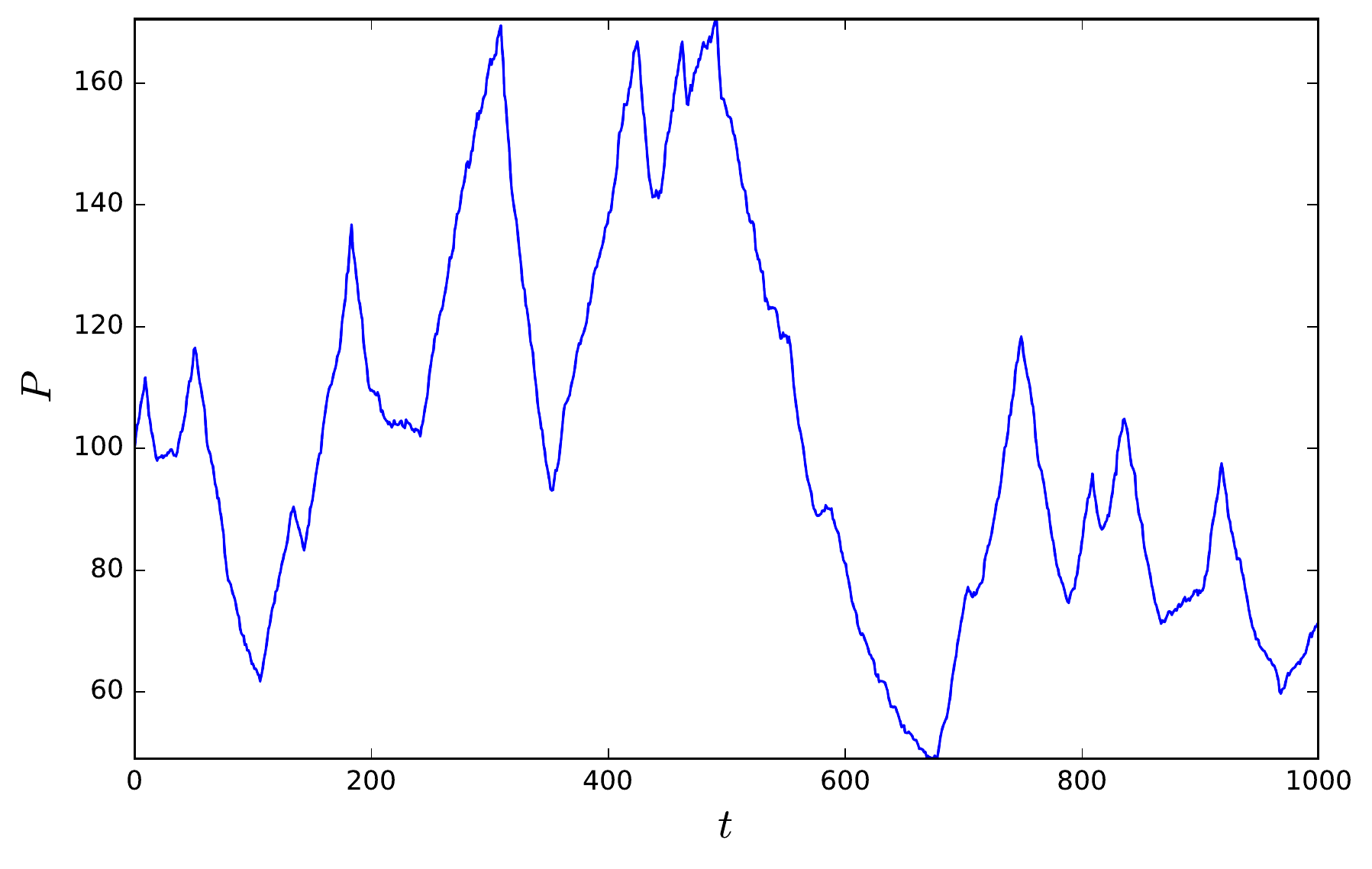}
 \caption{ \label{x1plot} Stock time series for the simulation
 in section \ref{sub_sec:simulation}. }
\end{center}
\end{figure}

Consider the system of $N=1000$ agents trading over the period of $T=1000$ steps.
At each time step each agent decides to participate in a trade with probability 
$\rho=1/10$. If it participates in a trade, it will choose the buy side with the 
probability $p_b$, and the sell side with the probability $p_s=1-p_b$,
such that $p_b/p_s=e^{\psi(t)}$. Regardless of the side of the trade
the agent will submit a limit price for its order chosen as a gaussian
draw around the most recent stock price with the standard deviation
being a stationary volatility sentiment process
$\sigma={\cal N}(0.02,0.005)$.
At the beginning we give each agent $1000$ shares of stock at the initial
price $P_1=100$, and ${\cal N}(10^5,10^3)$ units of cash. 

The sentiment process $\psi(t)$, governing the buy/sell decision
making of the agents, is a Markov chain, with three possible states,
and the corresponding equilibrium prices (\ref{flow_balance}) being
\begin{itemize}
\item
{\it Buy} with sentiment $\psi_b=1$, and equilibrium price $P_e^{(b)}=271$.
\item
{\it Neutral} with sentiment $\psi_n=0$, and equilibrium price $P_e^{(n)}=100$ .
\item
{\it Sell} with sentiment $\psi_s=-1$, and equilibrium price $P_e^{(s)}=36$.
\end{itemize}
The Markov chain $\psi(t)$ will be initialized randomly at $t=1$,
and the switch between the different sentiment states will occur according to the transition probabilities matrix
\begin{equation}
\label{general_buy_sell}
||a_{ij}||=\begin{pmatrix}P_{b,b} & P_{b,n} & P_{b,s} \\
P_{n,b} & P_{n,n} & P_{n,s} \\
P_{s,b} & P_{s,n} & P_{s,s} \\
\end{pmatrix},
\end{equation} 
where $P_{i,j}$ is  the probability of moving from 
$i$ to $j$ in one time step. In order to realize a long-lived
hidden states we will be considering the values of $a_{ij}$
drawn uniformly from the interval
\begin{equation}
a_{ii}\in [0.95,0.98]\,.\label{a_constraint}
\end{equation}
Then due to (\ref{a_constraint}) the system will spend on average 20-50
time steps in each sentiment state.
In such a manner we will run multiple trading simulations,
drawing the probability matrix $a_{ij}$ randomly,
taking into account the constraint (\ref{a_constraint}),
and generate the corresponding stock price time series.

\subsection{Single HMM reconstruction}\label{sub_sec:reconstruction}

In the previous subsection we discussed the general market environment
which will be used in this section to generated the stock price
time series governed by a buy/sell sentiment process of the Markov
chain kind.
In this section we are going to reconstruct the hidden transition matrix, $a^{{\rm fit}}_{ij}$,
from the observed stock price behavior using the Baum-Welch algorithm
of the HMM. We will begin by running a single simulation and explaining
our approach, and in the next subsection we proceed to running multiple simulation and gathering
fit results.
%The trained hidden transition matrix shall be the determination of the original hidden transition matrix (\textit{i.e.} $a^{trained}_{i j}\approx a_{i j}$). 

While we use the methods of the HMM to fit the sentiment transition probabilities matrix $a_{ij}$,
the stock market environment discussed above is not a Hidden Markov system.
Indeed, while the hidden sentiment process is constructed to be
of the Markov chain type, the observed stock price is derived from an intersection
of supply and demand at each time step, rather than obtained using an emission
probabilities matrix (as in the HMM) from the hidden states to the observed states.

Specifically, in the HMM for each hidden state the emission probabilities for
observed states are always the same. But in the stock market simulation
each sentiment does not uniquely determine distribution of possible
stock prices. In fact, the price is expected to be essentially influenced by the recent
sentiments, for instance, it matters whether the sentiment has recently
decreased or increased. 
Therefore a more appropriate way to fit
the sentiment time series would involve a method allowing to keep track of the
sequence of states, rather than one state. We will employ such a method
in the next section, where we will be considering application of the Recurrent
Neural Networks to the problem of inferring the sentiments.

%\subsection{Pre-defined transition matrix}\label{definedtransitionmatrix}

Let us incorporate a buy/sell sentiment driving process of the Markov chain type with the
transition probability matrix
\begin{equation}
\label{a_matrix_section_3}
a_{ij}=\begin{pmatrix}0.95 & 0.025 & 0.025 \\
0.035 & 0.93 & 0.035 \\
0.05 & 0.05 & 0.9 \\
\end{pmatrix}.
\end{equation} 
into the market system described above in this section. Running a simulation 
over $T=1000$ steps we have generated the stock price time series
plotted in figure \ref{x1plot}. We are going to take the stock price time series produced
in this simulation as an observable input. 
Following the Baum-Welch algorithm
we were able to reconstruct the transition probabilities matrix as
\begin{equation}
a^{{\rm fit}}_{ij}=\begin{pmatrix}0.93\ & 0.07 & 0\ \\
0.06 & 0.82 & 0.12 \\
0 & 0.09 & 0.91 \\
\end{pmatrix}.
\end{equation} 
Notice that this result is rather close to the actual transition
probabilities matrix (\ref{a_matrix_section_3}).
It is notable that such an accuracy drops sharply if one violates the constrain given in (\ref{a_constraint}). The reason for taking that constraint is to have the market reside in a state with the given
sentiment for a long enough time. This way the stock price can spend enough time close
to the corresponding equilibrium value $P^{(e)}$, thereby allowing the HMM
to train on that value as the one corresponding to each specific sentiment $\psi$.
On the other hand in a system with short-lived sentiment states
assumptions of the HMM will be invalid, as discussed above, rendering the Baum-Welch
algorithm inapplicable.
We will repeat such a simulation and reconstruction below multiple
times and confirm this result.

Applying the Viterbi algorithm we obtained the fit score $0.4$,
that is, only $40\%$ of the Viterbi predictions of the underlying sentiments match the reality.
This is almost as good as a random guess of one out of three sentiments,
as we will confirm below by running multiple simulations and aggregating
scores of the Viterbi predictions. Such a poor performance of the Viterbi
algorithm for the market sentiment reconstruction is in fact anticipated, and as discussed
above can be explained
by the fact that the stock market simulation is not really a Hidden Markov Model.

\subsection{Repeated HMM reconstructions}
\label{repeated_hmm}

In this subsection we are going to discuss the results of the repeated simulations
of the kind described in subsection \ref{sub_sec:reconstruction}. For each simulation
we are going to initialize the transition probabilities matrix of the hidden sentiment
Markov process randomly, with the only demand that it satisfies the constraint (\ref{a_constraint}).
For each generated sentiment process we simulate the stock market evolution
in the setup described in subsection \ref{sub_sec:simulation}.
We then perform the Baum-Welch fit for the transition probabilities matrix $a_{ij}$,
and the Viterbi fit for the sentiment states $\psi(t)$.

\begin{figure}
\begin{center}
 \includegraphics[width=15.5cm, height=11cm]{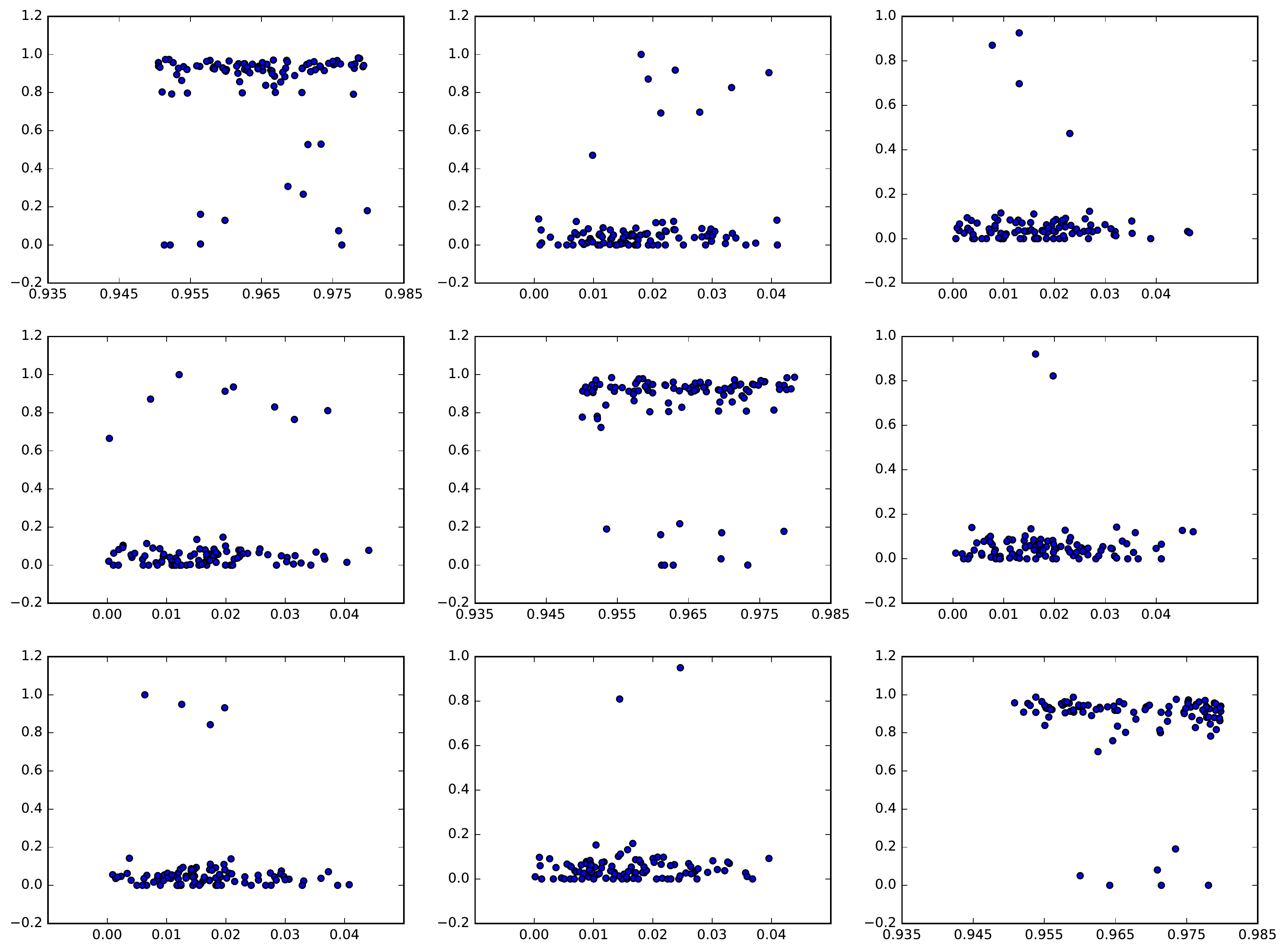}
 \caption{ \label{multiplefit}
Results for the sentiment transition probabilities matrix fit in subsection \ref{repeated_hmm} over 100 simulations.
 Each pane of the $3\times 3$ plot
corresponds to the fit results for one of the nine matrix elements.
The horizontal axis represents
the actual value of the matrix element $a_{ij}$, and the vertical axis represents the Baum-Welch fit $a_{ij}^{{\rm fit}}$.}
\end{center}
\end{figure}

The result of the Baum-Welch fit is a $3\times 3$ plot in figure \ref{multiplefit}, where each subplot corresponds to
a hidden transition probabilities matrix element.
Specifically, on each pane we provide a scatterplot of the Baum-Welch
fit for the transition probability matrix elements, $a_{ij}^{{\rm fit}}$, vs.
the actual transition probability matrix elements, $a_{ij}$, used in that simulation.

\begin{figure}
\begin{center}
 \includegraphics[width=12.8cm, height=8cm]{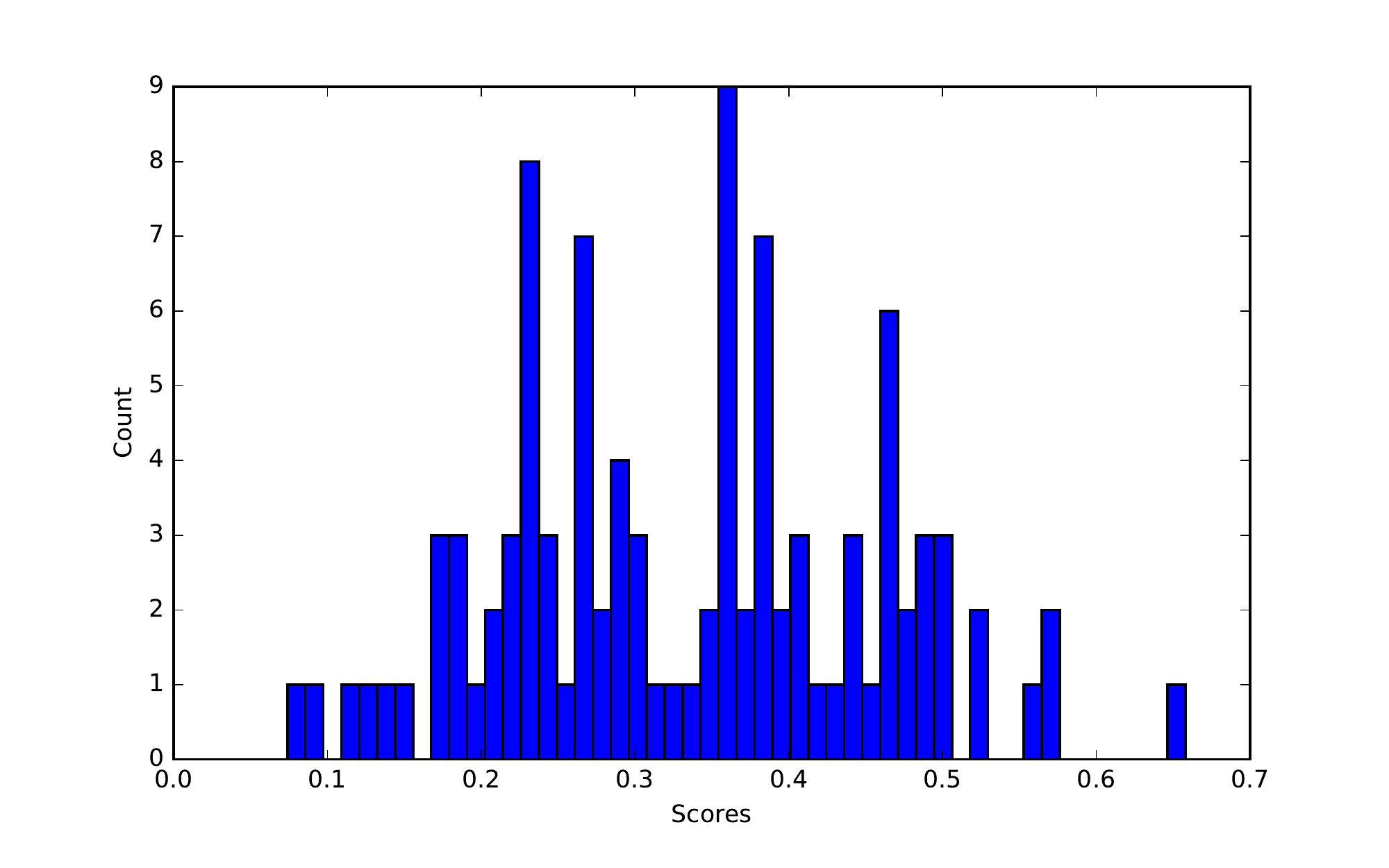}
 \caption{ \label{virebifit}
 Results for the sentiment states fit in subsection \ref{repeated_hmm} over 100 simulations,
 obtained using the Viterbi algorithm. The score on the $x$-axis represents the fraction of states
 reconstructed correctly.}
\end{center}
\end{figure}

For each simulation described above we also use the Viterbi algorithm to
reconstruct the sentiment state $\psi(t)$ at each time step $t$.
We plot the results in figure \ref{virebifit}. The mean score is $0.33$,
which is as good as guessing one of three sentiment states at random.
This is a reflection of the fact that the sentiment market environment,
as discussed above, is not a hidden Markov system.

\section{Sentiment fit using Recurrent Neural Network}
\label{Sentiment_fit_using_Recurrent_Neural_Network}

In section \ref{section_sentiment_hmm_reconstruction} we discussed an approach to reconstruct properties
of the hidden sentiment process from the observed stock price behavior using the
methodology of the HMM. We have demonstrated that the Baum-Welch algorithm of the HMM can be used
to recover the transition probabilities matrix for the sentiment Markov process
with a sufficiently long-lived sentiment states.
We have also applied the Viterbi algorithm to estimate the underlying sentiment states
from the observed stock prices and demonstrated that the resulting predictions are not satisfactory,
and are no better than a random guess. We have argued
in subsection \ref{sub_sec:reconstruction}
that such a poor performance of the HMM fit for the sentiment states
is due to the fact that the simulated stock market, unlike the HMM, is
influenced by the sequence of recent states, rather than a single current state.
As an alternative to the Viterbi algorithm in this section we attempt another approach, using the Recurrent Neural Networks,
to the problem of recovering sentiment
states from the stock price time series.  We refer reader to \ref{sec:rnn} for a review of the RNN,
relevant for the purposes of this paper.

\begin{figure}
\begin{center}
 \includegraphics[width=12.8cm, height=8cm]{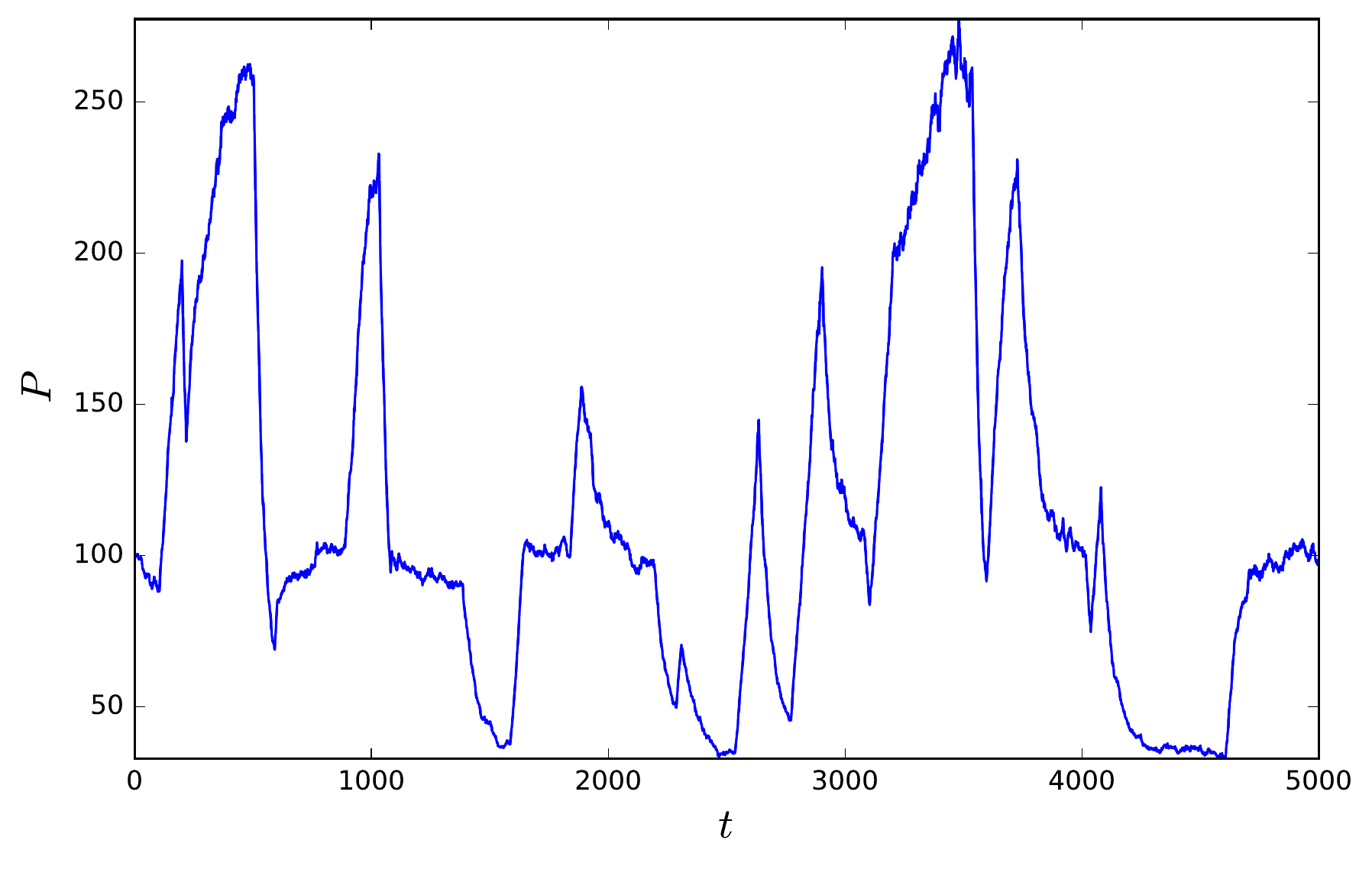}
 \caption{ \label{stocks_section_5} Stock time series for the simulation
 in section \ref{Sentiment_fit_using_Recurrent_Neural_Network}.
}
\end{center}
\end{figure}

An RNN is a neural network designed to fit time series, and capable of remembering and training on sequences
of states.
An input for an RNN is a time series $\{P_t\}$, $t=1,\dots$ of an arbitrary length,
and one can train RNN on sequences of data of length $T$ (unfold the RNN
into $T$ time layers). We are interested
in reading an output $\hat\psi_t$ at each $t$, which will be the buy/sell sentiment of the market 
at time $t$. The hat indicates that this is the fit, rather than the true sentiment $\psi_t$.

Consider the system of $N=1000$ agents over $T=5000$ simulation steps, driven by a buy/sell sentiment
process of the Markov chain type with the states $\psi=-1,0,1$, and the transition probabilities matrix
of the form (\ref{general_buy_sell}) defined to be equal to 
\begin{equation}
\label{a_matrix_section_4}
a_{ij}=\begin{pmatrix}0.9948 & 0.0002 & 0.005 \\
0.0016 & 0.9962 & 0.0022 \\
0.0044 & 0.0025 & 0.9931 \\
\end{pmatrix}.
\end{equation} 
We also include the volatility sentiment as a stationary process $\sigma(t)\sim {\cal N}(0.02,0.005)$.
Running the simulation we obtain the stock price time series in figure \ref{stocks_section_5}.
Notice that the system spends about 100 steps in each sentiment state.
We split the $5000$ steps into the first $4500$ steps, used as a train set,
and the last $500$ steps, used as a test set. We train the RNN on the chunks of data with $T=100$ steps
from the train set.
The length of the memory layer is taken to be $200$.
We then ask the RNN to predict sentiments from the stock price observations in the test set.
The train score then turns out to be $0.51$, and the test
score is $0.54$, reflecting the fractions of sentiment states inferred correctly by the RNN.
This is significantly better than $0.33$ score of the random guess.

\begin{figure}
\begin{center}
 \includegraphics[width=7.5cm, height=5cm]{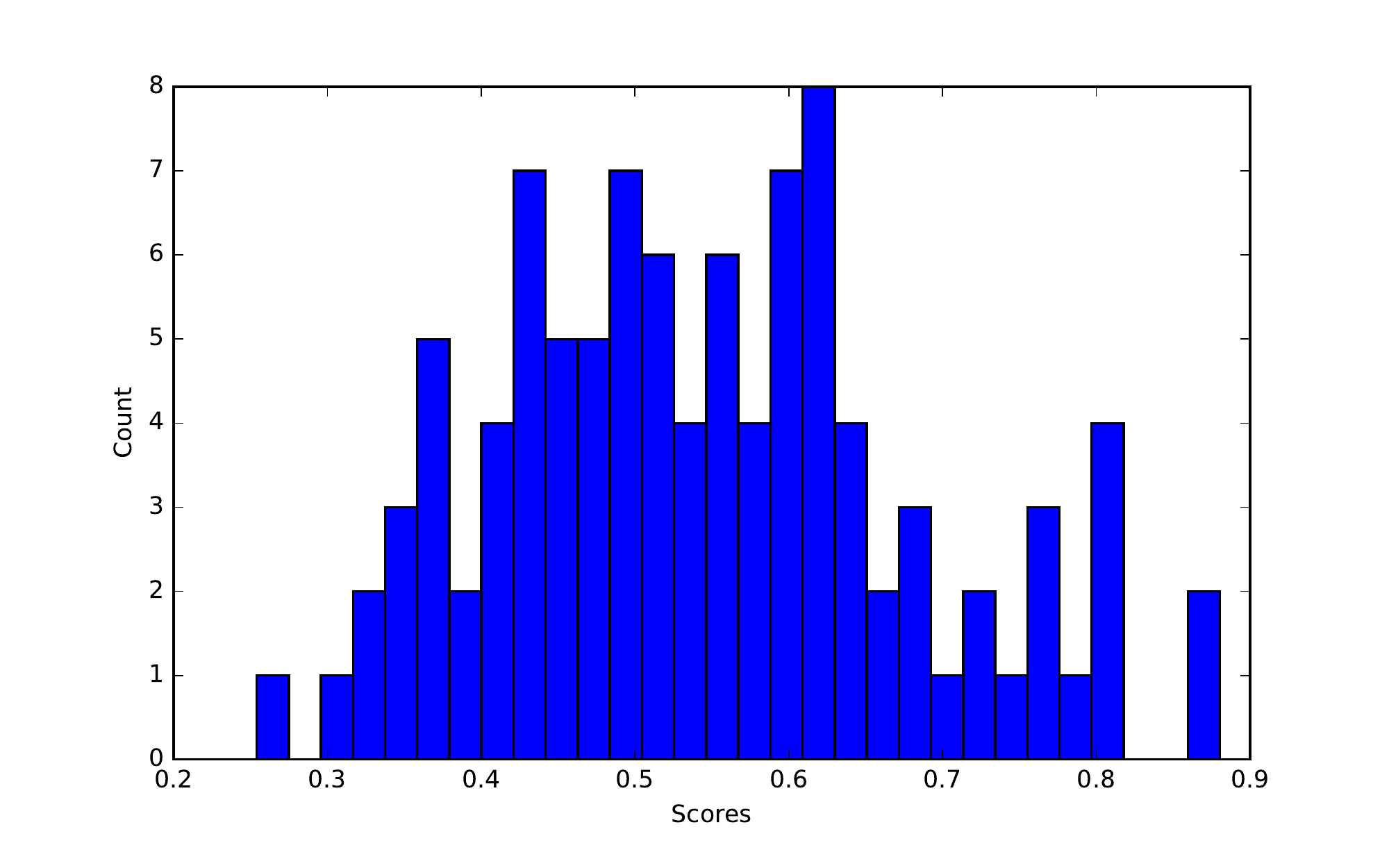}
  \includegraphics[width=7.5cm, height=5cm]{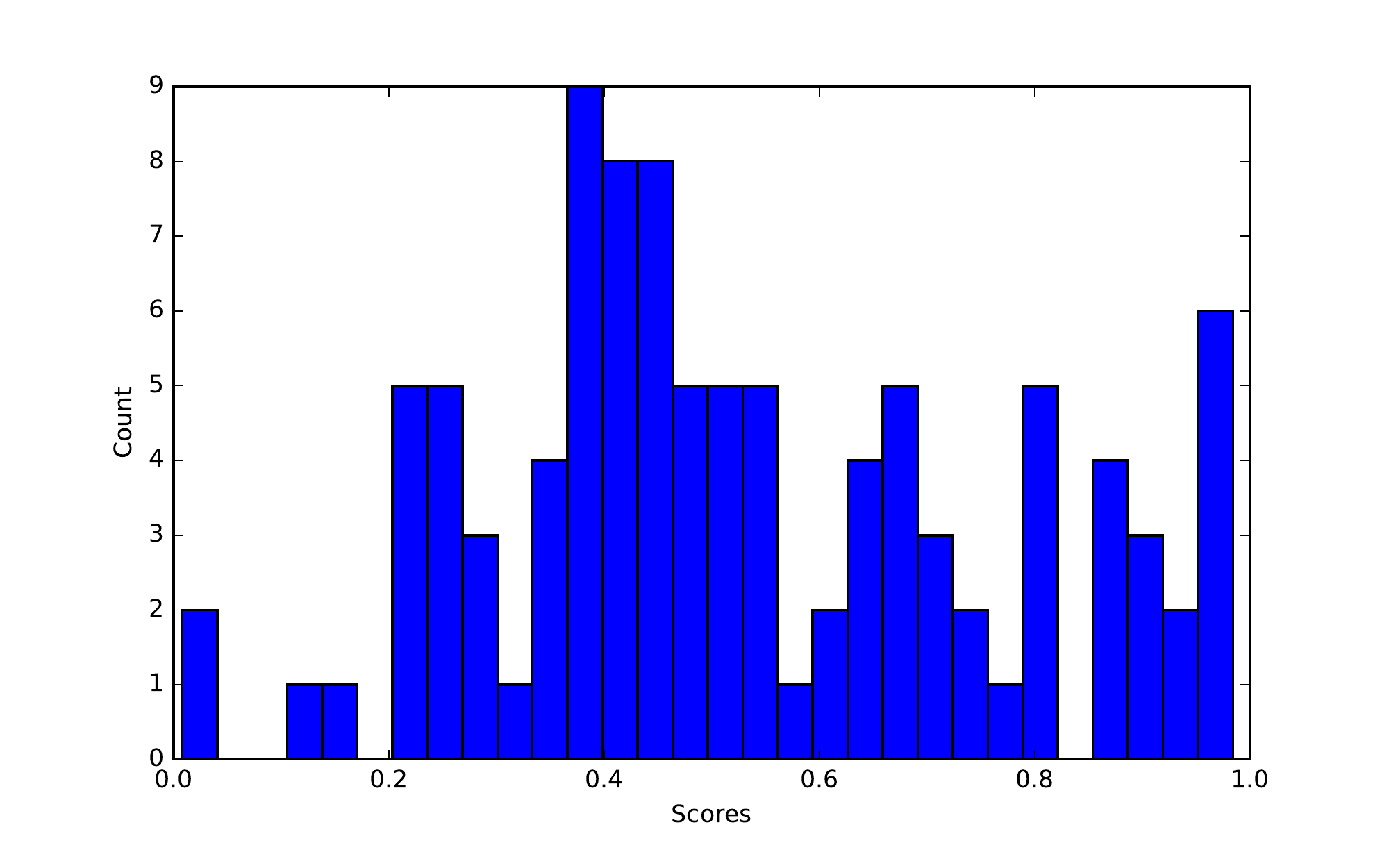}
 \caption{ \label{rnnfit}
 Results for the sentiment states RNN fit in section \ref{Sentiment_fit_using_Recurrent_Neural_Network} for 100 simulations. The score on the $x$-axis represents the fraction of states
 reconstructed correctly.
 The train set is the first $90\%$ of the data points, the test set is the last $10\%$.
 The mean of the train set (on the left) is $0.55$, the mean of the test set (on the right) is $0.53$. }
\end{center}
\end{figure}

In order to confirm this result
we repeat the calculation over 100 simulations, drawing the diagonal elements of the
sentiment transition probability matrix uniformly from the interval $[0.97,1)$.
We noticed that in order to optimize the average RNN performance one should unfold the RNN for
the $T$ steps in the range $T\simeq [25,75]$. In figure \ref{rnnfit} we plot the scores obtained for $T=50$.
This makes sense, because we need $T$ to be larger
than the length of the transition period between the sentiment states.
On the other hand $T$ should be smaller than the life-time of a sentiment state.
The train set score and the test set score are equal to $0.55$, $0.53$ respectively,
which is an improvement compared to the 0.33 random guess score. We speculate that not quite
perfect performance of an RNN is due to the properties of the simulated stock price time series.
It would be interesting to classify systematically what kind of simulated stock price times series
in a sentiment-driven stock market framework allows for a more accurate reconstruction of the underlying
sentiment states.

\section{Conclusions and discussion}
\label{sec:discussion}

In this paper we discussed the problem of inferring the properties
of the hidden sentiment process from the observed stock price behavior
in a sentiment-driven simulated stock market framework of \cite{Goykhman2017}.
We have mostly considered the systems in which the sentiment process
is a Markov chain, and approached the problem of retrieving the Markov
transition probabilities matrix, and the sentiment states themselves,
from the stock price time series. To tackle this problem we proposed
to use the methods of the Hidden Markov Model and the Recurrent Neural Network.

We demonstrated that the Baum-Welch algorithm of the Hidden Markov Model
allows to successfully reproduce the sentiment transition probabilities
matrix, provided the system exhibits a long-lived sentiment states.
We also demonstrated that the prediction performance of the Viterbi
algorithm of the HMM, applied to infer the sentiment states at each time step, is as good as a random guess.
This is due to the fact that, the Viterbi algorithm is predicting the most probable path between hidden states and observation, assuming that for the given hidden state the distribution of observations is
always the same.
In other words, in the HMM the observable state, which in our case is the stock price,
is determined locally from the hidden state, which in our case is the sentiment,
by the fixed emission probability matrix. Therefore the distribution of observables
is always the same for the same hidden states.
However, in the case of the simulated
stock price the distribution of a price at the given time moment depends on the sequence
of the recent hidden sentiment state, not only on the single most recent sentiment.

%We have also applied the Viterbi algorithm to estimate the underlying sentiment
%states
%from the observed stock prices and demonstrated that the resulting predictions
%are not adequate. This is due to the fact that, the Viterbi algorithm is predicting the most probable path between hidden states and observation. That is essentially to predict the most probable emission matrix. In our model however, we studied the hidden state transition matrix and the corresponding trained probability matrix, and not the emission paths. Thus, the results of  the Viterbi algorithm shown no particular preference in choosing a path from hidden states to observed ones.

We observed that in order to be able to recreate the underlying
sentiment state from the observed stock price time series
we need to have a model which will fit the current stock price against the sequence
of the recent sentiment states, not just against the current sentiment state.
For this purpose we proposed to use the technology of the Recurrent Neural Network,
which is known as a framework capable of remembering the sequence of states
and fitting time series.
We demonstrated that the accuracy of the RNN method is around $50\%$,
being significantly better than the random score of $33\%$.
We suggest the further tuning of the RNN can improve the fit score even further. 

%This can be due to the time interval $T$, which corresponds to the trained RNN. Changing this interval can be studied in the future for optimisation. 

In this paper we have refrained from applying the sentiment-driven
market framework to the real-world market, leaving this interesting
direction for future work. We note that
our results for the Baum-Welch and RNN performance suggest that
our method can be adequate for day rather than intra-day 
stock price time series, so that one would have regimes with potentially
long-lived sentiment states.
%the more the hidden states (buy, sell, neutral) are long lived the better one can reproduce the transition probabilities and their corresponding fits. This suggests that our method can be adequate for long positions and not daily or intra-day trading.
It is worth mentioning that our approach can be used for more complicated models involving option trading. This is because most of options are traded over a long run. We leave studying such models for a future work.  

\section*{Acknowledgements} M.G. was supported by Oehme Fellowship.

%\clearpage

\appendix
\section{Hidden Markov Models}\label{sec:hmm}

%In this paper we discuss the problem of reconstruction of
%the underlying sentiment processes governing a stock price dynamics in a sentiment-driven stock
%market simulation framework. An example of such a problem 
%would be recreating the sentiment $\psi(t)$ resulting in the
%stock time series observed in figure \ref{stocks_section_2},
%originated in the model descrived in section \ref{sec:buy_sell_sentiment}. 
%In that case it is clear that there are three well separated regimes, defined by the values
%of the sentiment, and we can reconstruct those values using 
%the average cash flow balance equation (\ref{flow_balance})
%in each of those regimes.

%Most of the times we expect the problem of reconstructing the
%properties of the underlying market sentiments to be far less
%trivial.

%In this paper we are interested in a type of a sentiment-driven markets
%in which the sentiment process is of a stationary Markov chain type.
%The problem is then to determine the transition
%probability matrix of the sentiment Markov chain, and infer
%the most likely underlying sentiment states from the 
%observed stock price time series.
%One way to address this problem is to apply the methodology of the
%Hidden Markov Models. Therefore

In this appendix we
review the known techniques of the Hidden Markov Models. For an excellent
introduction we refer the reader to \cite{Rabiner1989}.

The framework of HMM is based on the concept of an ordinary Markov chain.
Consider a random process $\{x_t\}$ in discrete time, $t=1,2,\dots,T$,
where $T$ is the length of the period over which we study the process.
At each time step $t$ the random variable $x_t$ can 
be in one of $N$ states, denoted by $X_1, X_2, \cdots, X_N$.
We cannot observe the states $\{x_t\}$
directly. Instead we observe the process $\{y_t\}$,
where each $y_t$ can be in one of $M$ states $Y_1$, $Y_2$,..., $Y_M$.

The HMM is characterized by the following parameters:
\begin{itemize}
\item Hidden state transition probability distribution $A=\{a_{ij}\}$, where $a_{ij}$ is the $N\times N $ matrix of transition probabilities between hidden states, \textit{i.e.}
\begin{displaymath}
a_{ij}=P(x_{t+1}=X_j|x_t=X_i), \qquad 1\leqslant i, j\leqslant N.
\end{displaymath}
\item Observable state probability distribution $B=\{b_{ik}\}$, where $\{b_{ik}\}$ is the $N\times M $ matrix of emission probabilities between hidden states to observed states,
\textit{i.e.}
\begin{displaymath}
b_{ik}=P(y_{t}=Y_j|x_t=X_{i}), \qquad 1\leqslant i\leqslant N,1\leqslant  k\leqslant M.
\end{displaymath}
\item Initial distribution $\pi=\{\pi_i\}$, where $\pi_i$ is $N$-tuple of probabilities of the initial hidden states, \textit{i.e.}
\begin{displaymath}
\pi_i=P(x_{1}=X_{i}),\qquad 1\leqslant i\leqslant N.
\end{displaymath}
In this manner we shall define the discrete HMM by a triplet: 
\begin{equation}
\lambda=(A, B, \pi).
\end{equation}
\end{itemize}  

If we observe the sequence $\{y_t\}$, $t=1,\dots,T$,
and we know the spectrum of observable states $\{Y_t\}$,
and the dimension of hidden states, we can use the Baum-Welch
algorithm to infer the parameters $\lambda$ of the HMM.
We begin by initializing $\lambda$ by a guess, and then
iterating the following procedure until it converges.
First we calculate the forward and backward probabilities
\begin{align}
\hat\alpha_{it}&=P(x_t=X_i|y_1\cdots y_t)\,,\qquad \hat\beta_{it}=\frac{1}{c_{t+1}\cdots c_T}\,
P(y_{t+1}\cdots y_T|x_T=X_i)\,,\\
c_t&=P(y_t|y_1\cdots y_{t-1})
\end{align}
using the the recursion relations
\begin{align}
\alpha_{i1}&=\pi_i\,b_{ik_1}\,,\quad c_{t+1}\,\hat\alpha_{i\,t+1}
=\sum_j\hat\alpha_{jt}\,a_{ji}\,b_{i\,k_{t+1}}\,,\\
\hat\beta_{iT}&=1\,,\quad c_{t+1}\,\hat\beta_{it}=\sum_j\hat\beta_{j\,t+1}\,a_{ij}\,b_{j\,k_{t+1}}\,,\\
\sum_{i}\hat\alpha_{it}&=1\,,\quad \hat\alpha_{i1}=\frac{\alpha_{i1}}{c_1}\,,
\end{align}
where we have denoted $y_{t}=Y_{k_t}$,
Next we calculate the probabilities.
\begin{align}
\gamma_{it}&=P(x_t=X_i|y_1\cdots y_T)=\hat\alpha_{it}\,\hat\beta_{it}\,,\\
\xi_{tij}&=P(x_t=X_i,x_{t+1}=X_j|y_1\cdots y_T)=\frac{1}{c_{t+1}}\,\hat\alpha_{it}\hat\beta_{j\,t+1}\,a_{ij}\,b_{j\,k_{t+1}}\,.
\end{align}
Finally we update the $\lambda$ parameters as 
\begin{equation}
\pi_i=\gamma_{i1}\,,\qquad a_{ij}=\frac{\sum_{t=1}^{T-1}\xi_{tij}}{\sum_{t=1}^{T-1}\gamma_{it}}\,,
\quad  b_{ik}=\frac{\sum_{t=1}^T\delta(y_t=Y_k)\,\gamma_{it}}{\sum_{t=1}^T\gamma_{it}}\,.
\end{equation}

The underlying hidden states $\{x_t\}$ can be inferred using the Viterbi
algorithm. One defines two auxiliary matrices $R_{it}$, $Q_{it}$
of size $N\times T$, where $R_{it}$ is the probability of the most
likely sequence $x_1\cdots x_t$ given the observed sequence $y_1\cdots y_t$,
such that $x_t=X_i$. It can be calculated recursively as
\begin{equation}
R_{i1}=\pi_i\,b_{ik_1}\,,\quad R_{it}=\max_j(R_{j\,t-1}\,a_{ji})\,b_{ik_1}\,.
\end{equation}
The $Q_{it}$ is $x_{t-1}$ of the most likely hidden sequence $x_1\cdots x_t$
given the observation $y_1\cdots y_t$. It can be calculated 
from the known $R_{it}$ as
\begin{equation}
Q_{it}={\rm arg\, max}_j(R_{j\,t-1}\,a_{ji})\,.
\end{equation}
The most likely hidden state at $t=T$ is then
\begin{equation}
x_T={\rm arg\, max}_i(R_{iT})\,,
\end{equation}
which is then used to initialize iterative calculation
\begin{equation}
x_t=Q_{x_{t+1}\,t+1}\,.
\end{equation}

\section{Recurrent Neural Network}
\label{sec:rnn}

In this appendix we
review the simplest kind of a Recurrent Neural Network (RNN).
An RNN is the neural network able to fit the data
represented as a time series. A typical problem which can be addressed by an
RNN is to predict the output times-series $\{y_t\}$ from the input
time series $\{x_t\}$, where $t=1,2,\dots$. An RNN is designed
to be flexible to the specific length of the time series.
We can train it on the chunks of data of the length $T$.
Suppose an input at the given time $t$ is a vector $x_t$ of dimension $S$ and an output $y_t$
is a vector of dimension $N$.
The $x_t$ is received by the input layer,
and the $y_t$ is produced by the output layer.
Similarly to the usual neural network
the RNN has a hidden layer, in between the input and the output layers.
The hidden layer $m_t$ is the `memory' of the RNN, influenced both by the current
input $x_t$, and by all the previous inputs $x_\tau$, $\tau=1,2,\dots,t-1$.
We represent the memory layer $m_t$ as a vector of size $M$.

An RNN is characterized by the hidden bias vector $b$,
output bias vector $e$, input-to-weight matrix $W$,
memory-to-memory matrix $V$, and memory-to-output matrix $U$.
We will describe these parameters collectively as
\begin{equation}
{\cal R}=(W,V,U,b,e)\,.
\end{equation}
The core functionality of the RNN is described by the equations
\begin{align}
m_t&=f\left(W\,x_t+V\,m_{t-1}+b\right)\,,\\
y_t&=U\,m_t+e\,,
\end{align}
with some classifier function $f$, which in this paper we choose to be $f\equiv\tanh$.
To summarize, the inner structure of the RNN is characterized by the $t$-independent
parameters ${\cal R}$ and by the $t$-dependent memory vectors $m_t$.
To train an RNN means to fit the parameters ${\cal R}$ to produce the desired
output $\{y_t\}$ from the given input $\{x_t\}$.

It is convenient to represent the desired outputs $y_t$ in such a way that at each
given time step $t$ we want the output $y_t$ to be equal to the vector $R_t$
with the components
\begin{equation}
R_{it}=\delta_{i\,r_t}\,,\quad i=1,\dots,N\,.
\end{equation}
Then we can represent the desired output as the time series of index values $\{r_t\}$
of the non-zero $y_t$ vector component.
We can match this desired output vector to the actual output $y_t$
by calculating the probabilities of the $r_t$ values
\begin{equation}
P_{it}=\frac{e^{y_{it}}}{\sum_j e^{y_{jt}}}\,,\quad i=1,\dots,N\,,
\end{equation}
and using those probabilities in the loss function, see, {\it e.g.}, \cite{Karpathy}
\begin{equation}
L=-\sum_t\log\left(\sum_i P_{it}\,R_{it}\right)\,.
\end{equation}

The RNN parameters ${\cal R}$ can then be tuned over many training iterations
using the backpropagation of the loss function $L$. We want the loss function
value to be as small as possible, so we will shift the values of the parameters ${\cal R}$
in the direction opposite to the gradients of the loss function $L$ w.r.t. those parameters.
We summarized all the key gradients here,
\begin{align}
\left[\nabla^{(y)}L\right]_{mt}&\equiv\frac{\partial L}{\partial y_{mt}}=P_{mt}-\delta_{m\,r_t}\,,\\
\left[\nabla^{(U)}L\right]_{ij}&\equiv \frac{\partial L}{\partial U^{ij}}
=\sum_t\left[\nabla^{(y)}L\right]_{it}m_{jt}\,,\\
\left[\nabla^{(e)}L\right]_{i}&\equiv \frac{\partial L}{\partial e^{i}}=
\sum_t\left[\nabla^{(y)}L\right]_{it}\,,\\
\left[\nabla^{(m)}L\right]_{it}&\equiv \frac{\partial L}{\partial m_{it}}=
\sum_j U^{ji}\left[\nabla^{(y)}L\right]_{jt}+\left[\nabla^{(m_n)}L\right]_{it}\,,\\
\left[\nabla^{(m_n)}L\right]_{it}&\equiv\sum_j\left[\hat\nabla^{(m)}L\right]_{j\,t+1}V^{ji}\,,\\
m_{it}&\equiv \tanh \hat m_{it}\,,\\
[\hat\nabla^{(m)}L]_{it}&\equiv \frac{\partial L}{\partial\hat m_{it}}=\left[\nabla^{(m)}L\right]_{it}
\,(1-m_{it}^2)\,,\\
\left[\nabla^{(W)}L\right]_{ij}&\equiv \frac{\partial L}{\partial W^{ij}} =
\sum_t[\hat\nabla^{(m)}L]_{it}\,x_{jt}\,,\\
\left[\nabla^{(V)}L\right]_{ij}&\equiv \frac{\partial L}{\partial V^{ij}} =
\sum_t[\hat\nabla^{(m)}L]_{it}\,m_{j\,t-1}\,,\\
\left[\nabla^{(b)}L\right]_{ij}&\equiv \frac{\partial L}{\partial b^{i}} =
\sum_t[\hat\nabla^{(m)}L]_{it}\,.
\end{align}

Once we know the gradients we know by how much we need to shift
the parameters ${\cal R}$ in order to decrease the loss function,
\begin{equation}
{\cal R}\rightarrow {\cal R}-\frac{r}{\rho}\,[\nabla^{({\cal R})}L]\,,
\end{equation}
where the learning rate is $r\simeq 0.1$, and the adaptive learning rate \cite{Duchi2011} is
\begin{equation}
\rho=\sqrt{\sum_n [\nabla^{({\cal R})}L]^2}\,,
\end{equation}
and the sum is done over all the iterations we have performed up till
now, inclusive.

\end{document}